\documentclass{nature}

\usepackage{amsmath}
\usepackage[pdftex]{graphicx}
\usepackage{dcolumn}  
\usepackage{bm}           
\usepackage{dsfont}
\usepackage{color}
\usepackage{braket}
\usepackage{booktabs}
\usepackage{hyperref}

\newcounter{defcounter}
\usepackage{amsthm}

\title{Excitonic Energy Transfer in Red Algal Photosystem I Reveals an Evolutionary Bridge between Cyanobacteria and Plants} 

\author{Mengyuan Cui$^{1,*}$, Zihui Liu$^{1,*}$, Miriam Izzo$^{2,*}$, Junhua Zhou$^{1,*}$, Enhu He$^{1}$, Vandana Tiwari$^{3,4}$, Petar H. Lambrev$^{5}$, R. J. Dwayne Miller$^{6}$, Joanna Kargul$^{2}$, Fulu Zheng$^{1}$, Ajay Jha$^{7,8}$, Hong-Guang Duan$^{1}$} 

\begin{document} 

\maketitle 

\begin{affiliations} 
\item Department of Physics, School of Physical Science and Technology, Ningbo University, Ningbo, 315211, P.R. China 
\item Solar Fuels Laboratory, Center of New Technologies, University of Warsaw, 02-097, Warsaw, Poland 
\item Stanford PULSE Institute, SLAC National Accelerator Laboratory, Menlo Park, California 94025, United States
\item Department of Chemical Science, Linac Coherent Light Source, SLAC National Accelerator Laboratory, Menlo Park, California 94025, United States 
\item Biological Research Centre, Szeged, Szeged 6726, Hungary
\item Departments of Chemistry and Physics, University of Toronto, 80 St George Street, Toronto, M5S 3H6, Ontario, Canada
\item Rosalind Franklin Institute, Harwell, Oxfordshire OX11 0QX, United Kingdom 
\item Department of Pharmacology, University of Oxford, Oxford, OX1 3QT United Kingdom  \\ 
\centerline{\underline{\date{\bf \today}}} 
\end{affiliations} 

\begin{abstract}

Photosystem I (PSI) converts light into chemical energy with near-unity quantum efficiency, yet its energy-transfer and charge-separation mechanisms remain debated. Evolution has diversified PSI architectures: cyanobacterial PSI trimers confine red-shifted pigments to the core, whereas plant PSI-Light Harvesting Complex I (PSI-LHCI) supercomplexes incorporate extensive peripheral red and charge-transfer states that reshape trapping. The unicellular red algae \textit{Cyanidioschyzon merolae} represents a key evolutionary intermediate, featuring a cyanobacterial-like monomeric core surrounded by three to five LHCR (light harvesting complex in red algae) subunits. This hybrid organization provides a unique system to bridge mechanistic models across lineages. We applied two-dimensional electronic spectroscopy (2DES) at ultralow temperatures (8 K and 80 K) to disentangle overlapping excitation pathways in \textit{C. merolae} PSI. Cryogenic measurements suppressed thermal broadening, resolving five dynamical components: sub-picosecond equilibration (0.3 - 0.8 ps) across the core-LHCR interface, subsequent population transfer (2.6 - 4 ps) into progressively lower-energy manifolds, and slower feeding (18 - 53 ps) into red pools distributed across both core and antenna. On the longest timescales (hundreds of ps to effectively infinite), a persistent ground-state bleach signifies excitons stabilised in terminal sinks. Notably, comparison of 8 K and 80 K spectra reveals that excitations are heterogeneously partitioned among multiple sinks at low disorder, whereas modest thermal activation (kT $\sim$ 55 cm$^{-1}$) promotes selective convergence into core-associated red chlorophylls. To interpret these dynamics, we employed atomistic excitonic Hamiltonians with time-nonlocal master equations, providing a quantitative framework for exciton migration and thermal redistribution. Together, these results demonstrate that {\em C. merolae} PSI broadens the kinetic funnel by distributing sinks across core and antenna, an evolutionary adaptation that extends spectral coverage whilst ensuring efficient trapping. These insights reconcile cyanobacterial and plant paradigms and illuminate how antenna expansion reshaped PSI function during the course of photosynthetic evolution.

\end{abstract}


Photosystem I (PSI) is a fundamental component of the light-dependent reactions of oxygenic photosynthesis \cite{Ref1, Ref2}. Present in cyanobacteria, algae, and plants, this multi-subunit pigment-protein supercomplex exhibits near-unity quantum efficiency in charge separation (CS) process after solar energy absorption \cite{Ref3, Ref4, Ref5, Ref6}. The extraordinary performance of PSI arises from its intricate network of pigment molecules, primarily chlorophyll (Chl) a, embedded within a protein scaffold that orchestrates the absorption, migration, and trapping of photo-excitation energy. These pigment arrays act cooperatively, mediating excitation energy transfer (EET) across distances of several nanometers with sub-picosecond to picosecond timescales, ultimately leading to primary CS in the reaction center (RC) and electron transfer to downstream cofactors \cite{Ref1, Ref7, Ref8, Ref9}. Despite this exceptional performance, the precise sequence of excitonic energy transfer, trapping, and charge separation remains a subject of debate. A unified picture has been elusive in part because PSI has diversified considerably during evolution, acquiring distinct antenna systems in cyanobacteria, red algae, and green algae/plants, as depicted in Fig.\ \ref{fig:Fig1}(a)\cite{RefPSI_e1, RefPSI_e2, Ref10}. Resolving how these evolutionary changes remodel the excitation landscape requires integrating structural data with ultrafast spectroscopic measurements. 

In cyanobacteria such as \textit{Synechocystis sp. PCC 6803} and \textit{Synechococcus sp. PCC 7002}, PSI typically assembles as a trimer, each monomer containing $\sim$95 chlorophyll a molecules and $\sim$20 carotenoids arranged within a conserved core\cite{Ref3, RefPSI_1}. A subset of low-energy ``red" chlorophylls absorb beyond 700 nm, forming energetic sinks that bottleneck transfer into the RC \cite{Ref12, Ref14, Ref19, Ref20, Ref28}. Two-dimensional electronic spectroscopy (2DES) studies of cyanobacterial PSI trimers revealed sub-100 fs equilibration among bulk core pigments, followed by ps-scale population of the red pools, which govern the overall trapping timescale\cite{Ref47}. Although the number and spectral position of these red pigments vary across species, their role as the dominant kinetic choke point is conserved. 

In land plants, PSI exists as a monomeric core surrounded by a belt of 8-10 LHCI (Lhca) heterodimers, roughly doubling the pigment count to $\sim$200 Chls per complex\cite{Ref5}. The PSI core in plants still hosts a few red Chls, but their spectral contribution is comparatively modest. The dominant red states in plant PSI arise from LHCI subunits (especially Lhca1/4, which harbor charge-transfer–like red states absorbing 730 - 740 nm)\cite{Ref16}. Thus, the plant PSI red pool is primarily antenna-based rather than core-based. 2DES studies of intact PSI-LHCI supercomplexes showed ultrafast ($<$0.5 ps) equilibration in the core, $\sim$3-4 ps transfer into LHCI red/CT states, and slower tens-of-ps components reflecting delayed release from these states\cite{Ref11, Ref13, Ref15, Ref17, Ref18, Ref46}. Thus, while the plant PSI expands spectral coverage, it also introduces more heterogeneous relaxation pathways, illustrating how antenna enlargement alters dynamics without compromising overall efficiency. 

The unicellular red algae \textit{Cyanidioschyzon merolae} occupy a pivotal evolutionary position between cyanobacteria and plants. Structural studies show that its PSI is monomeric, like in plants, but binds only 3 - 5 LHCR subunits (each with $\sim$11 - 13 Chls and multiple zeaxanthins), yielding an intermediate pigment count of $\sim$130 - 160 Chls\cite{RefPSI_2}. A distinctive feature is the incorporation of the eukaryotic subunit PsaO, which coordinates three Chls at the antenna-core interface. The core also retains cyanobacterial-like subunits such as PsaM, producing a hybrid organization. This architecture differs from both cyanobacteria and plants in two key respects. First, low-energy pigments are distributed across both the core and LHCR, creating dual sinks. Second, pigment bridges formed by PsaO and interfacial LHCR Chls enable equilibration across the core-antenna boundary, in contrast to the slower transfer observed in plants. Functionally, \textit{C. merolae} PSI thus represents an evolutionary intermediate, extending absorption through LHCR while maintaining a cyanobacterial-like trapping arrangement. Despite its significance, \textit{C. merolae} PSI has not been systematically explored using multidimensional spectroscopy. Most prior ultrafast work has focused on cyanobacterial cores or plant PSI-LHCI supercomplexes, with isolated Lhca subunits also studied at cryogenic temperatures\cite{Ref29, Ref47, Ref48}. At low temperatures, excited state red Chl population occurs in 4 - 6 ps, slower than at room temperature, reflecting a general deceleration of EET pathways in photosynthetic systems\cite{Ref30, Ref31, Ref32, Ref33, Ref34}. These studies highlighted the central role of low-energy pigments, whether core-localized red Chls, LHCI CT states, or both in shaping PSI kinetics. Yet the absence of a red algal 2DES study has left a critical gap between prokaryotic and eukaryotic paradigms.

Here, we address this gap by applying 2DES to PSI from \textit{C. merolae}. By combining ultralow-temperature 2DES measurements (8 and 80 K) complemented with dynamics calculations using time-nonlocal master equations and atomistic excitonic Hamiltonians, we resolve overlapping excitation pathways in PSI. This strategy reveals ultrafast equilibration across the core-antenna interface, picosecond population of low-energy states, and slower transfer into the RC. By directly comparing these dynamics with cyanobacterial and plant PSI, we demonstrate that \textit{C. merolae} PSI serves as a mechanistic and evolutionary bridge, revealing how incremental antenna expansion and redistribution of low-energy pigments reshaped energy transfer and trapping during the evolution of oxygenic photosynthesis. 


\section*{Results and Discussion} 

\begin{figure}[h!]
\begin{center}
\includegraphics[width=15.0cm]{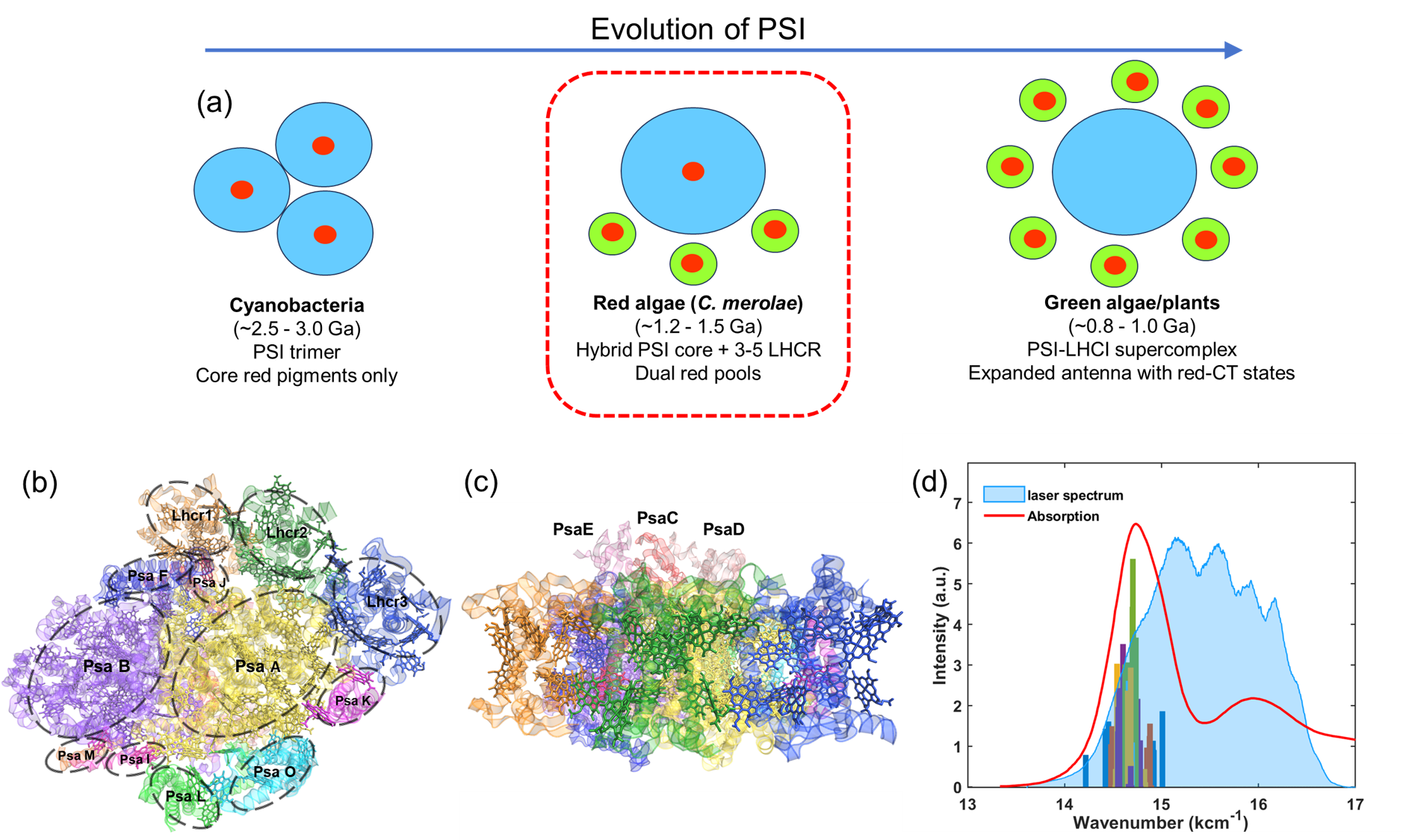} 
\caption{\label{fig:Fig1} Evolutionary context and spectral characterization of PSI. (a) Cartoon schematic illustrating the structural evolution of Photosystem I (PSI) from cyanobacterial trimers with core-localized red chlorophylls, through the red alga {\em C.\ merolae} intermediate featuring a monomeric core plus a small number of LHCR subunits representing dual red pools, to plant PSI-LHCI supercomplexes where the red pool is primarily antenna-based rather than core-based. (b) and (c) Top and side views of \textit{C.\ merolae} PSI, respectively, highlighting core subunits (PsaA/B) and peripheral Lhcr1-3, as well as bridging subunit PsaO. (d) Steady-state absorption spectrum (red line) of PSI compared with the excitation laser spectrum (blue area), showing broad pigment absorption with overlap to the experimental excitation bandwidth.} 
\end{center}
\end{figure}

The PSI-LHCR complexes were purified from genetically engineered \textit{C. merolae} strains expressing a His$_{6}$-tagged PsaD subunit, and their structural integrity, pigment composition, and photochemical activity were confirmed through biochemical characterization. Detailed methods and validation data, including absorption spectroscopy, SDS-PAGE analysis, and oxygen consumption assays, are presented in the Supplementary Information (SI) as Fig. S1 and S2. A structural overview of the PSI-LHCR complex is provided in Fig.\ \ref{fig:Fig1}(b) and (c), shown from the top and side views. The core RC subunits, PsaA and PsaB, which host the special pair of Chls responsible for primary charge separation, are highlighted at the center of the complex. Surrounding the RC are a series of small core subunits, including PsaF, PsaJ, and PsaK, which play critical roles in mediating energy transfer between the RC and the peripheral antenna. The LHCR antenna complexes: Lhcr1, Lhcr2, and Lhcr3 are positioned asymmetrically around the core, partially surrounding it and forming a modular light-harvesting belt. These subunits contribute additional pigments that extend the absorption cross-section of PSI and provide low-energy exciton states that are funneled toward the RC. 

The steady-state absorption spectrum of the \textit{C. merolae} PSI-LHCR complex is shown in  Fig.\ \ref{fig:Fig1}(d) (red solid line). As expected for red algal PSI, the spectrum exhibits distinct features characteristic of chlorophyll a, with prominent peaks centered near 630 nm and 680 nm. For comparison, the spectral profile of the excitation laser pulses employed in our 2DES measurements is overlaid as a blue shadow. The laser spectrum covers both chlorophyll absorption maxima, demonstrating that our experimental configuration allows us to simultaneously excite higher-energy pigments absorbing near 630 nm together with the bulk antenna states absorbing around 680 nm. This overlap ensures that the full excitonic manifold contributing to energy transfer in PSI is effectively accessed in our ultrafast measurements.

\subsection{Two-dimensional electronic spectroscopy of PSI complex} 

\begin{figure}[h!]
\begin{center}
\includegraphics[width=15.0cm]{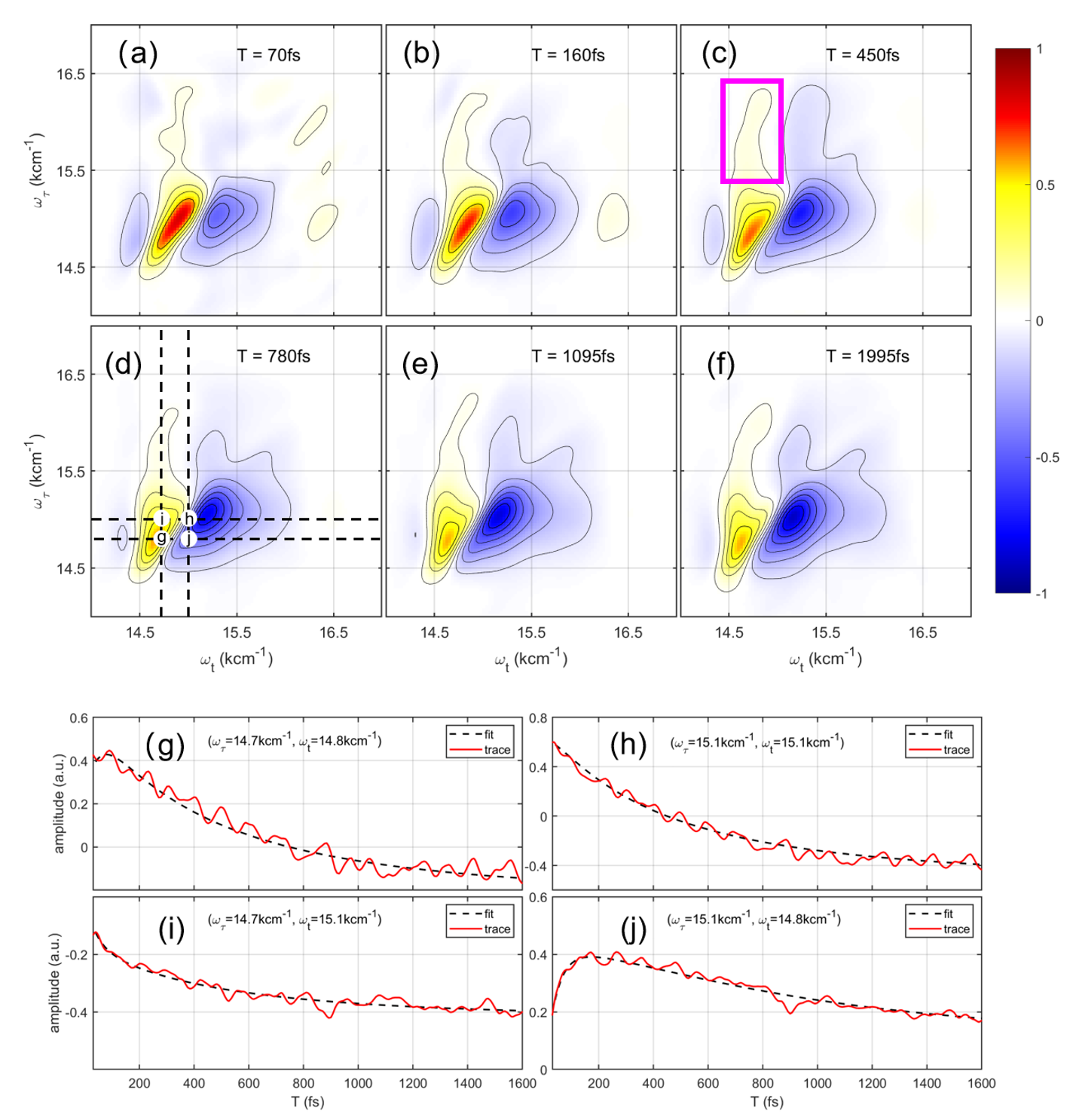}
\caption{\label{fig:Fig2} Two-dimensional electronic spectra of PSI at 8 K. Panels (a-f) show total real spectra at selected waiting times (70-1995 fs). Red (positive) features arise from ground-state bleach and stimulated emission, while blue (negative) features correspond to excited-state absorption. The persistence of overlapping red/blue peaks reflects strong excited-state absorption between single and double excited states, while the emergence of cross-peaks (magenta boxes) evidences downhill transfer pathways. (g-j) Time-dependent traces of selected diagonal and cross peaks (red) compared with global fits (black dashed), validating the kinetic model used to extract transfer timescales. }
\end{center}
\end{figure}

The 2DES has emerged as a unique probe to disentangle different excited state relaxation pathways, offering simultaneous femtosecond temporal resolution and high spectral selectivity \cite{Ref35, Ref36, Ref37, Ref38, Ref39, Ref41, Ref42, Ref44}. Unlike linear absorption or pump-probe spectroscopy, 2DES separates excitation and detection frequencies, enabling the direct observation of energy transfer events via off-diagonal cross-peaks and allowing the tracking of coherent phenomena through oscillatory modulations in signal amplitudes \cite{Ref44, Ref45}. Following steady-state characterization, we performed 2DES on the \textit{C. merolae} PSI-LHCR complex at 8 K, where suppressed thermal motion sharpens transitions and enhances the visibility of cross-peaks. Representative real-part 2D spectra at waiting times (T) of 70, 160, 450, 780, 1095, and 1995 fs are shown in Fig.\ \ref{fig:Fig2}(a) to (f). In these spectra, positive signals (red) correspond to ground-state bleaching (GSB) and stimulated emission (SE), whereas negative signals (blue) reflect excited-state absorption (ESA). Measurements at T = 0 fs were omitted because of pulse overlap artifacts that prevented reliable separation of absorptive and dispersive components. At T = 70 fs (Fig.\ \ref{fig:Fig2}(a)), a strong composite feature is observed near excitation $\sim$14,500 cm$^{-1}$ and detection $\sim$16,500 cm$^{-1}$. Here, positive GSB/SE overlaps with negative ESA, reducing the apparent bleach amplitude. This overlap reflects strong coupling between pigments, where excitation of bulk chlorophylls feeding into higher-energy states simultaneously gives rise to induced absorption. Above $\sim$15,500 cm$^{-1}$ on the detection axis, the spectra display predominantly ESA, consistent with access to higher-lying excitonic manifolds. Notably, at this low temperature, several off-diagonal cross-peaks are clearly resolved (highlighted in Fig.\ \ref{fig:Fig2}(c)), indicating direct downhill transfer channels from bulk antenna pigments into lower-energy states, including the red pools associated with both LHCR and core pigments. By 160 fs ((Fig.\ \ref{fig:Fig2}(b)), the overall line shape remains dominated by overlapped red-blue features, though the ESA contribution increases in relative strength. This suggests progressive population of excited states that participate in absorption from already occupied levels. The coexistence of GSB and ESA persists through 2 ps (Fig.\ \ref{fig:Fig2}(a) to (f)), indicating that low-energy excitations are populated early but remain dynamically coupled to higher-energy states, a hallmark of strongly delocalized excitonic networks. 

To probe these dynamics, we extracted time-domain traces from two diagonal and two cross-peaks marked in Fig.\ \ref{fig:Fig2}(d). Diagonal peaks monitor excitations that remain on initially excited bulk states, whereas below-diagonal cross-peaks report population transfer to lower-energy acceptors. For example, the feature linking excitation near 15,000 cm$^{-1}$ with detection at $\sim$14,700 cm$^{-1}$ (Fig.\ \ref{fig:Fig2}(g)) grows within the first few hundred femtoseconds, consistent with rapid downhill transfer from bulk antenna pigments to red-shifted sinks. The diagonal trace at $\sim$15,100 cm$^{-1}$ (Fig.\ \ref{fig:Fig2}(h)) shows corresponding decay, reflecting depletion of high-energy populations. Together, these pairs of diagonal and cross-peak dynamics provide direct spectroscopic evidence for population flow from bulk chlorophylls to progressively redder states. To quantify these behaviors, we employed a global fitting approach \cite{Valentyn GFP} (see SI). This analysis applies a common set of exponential components across all traces, yielding fits (black dashed lines in Fig.\ \ref{fig:Fig2}(g) to (j)) that closely reproduce the experimental kinetics. The excellent agreement confirms that the observed decays and rises can be captured by a shared set of dynamical processes, spanning sub-100 fs equilibration through multi-picosecond population of red pigments. While specific time constants are discussed later, at this stage the analysis demonstrates that the spectral signatures of bulk-to-red transfer are robust and reproducible across multiple probe regions. Extending the waiting time to hundreds of picoseconds (see SI, Fig.\ S8) reveals a marked simplification of the spectra. By $\sim$100 ps, the ESA contributions vanish entirely, leaving only positive bleach features corresponding to the lowest exciton states. This transition indicates that excitations have fully relaxed into long-lived red sinks, which persist as ground-state bleach signals after excited-state absorption pathways have decayed. The persistence of these GSB signals highlights the stability of the lowest-energy pigments, which act as terminal sites for energy funneling prior to charge separation.

\begin{figure}[h!]
\begin{center}
\includegraphics[width=15.0cm]{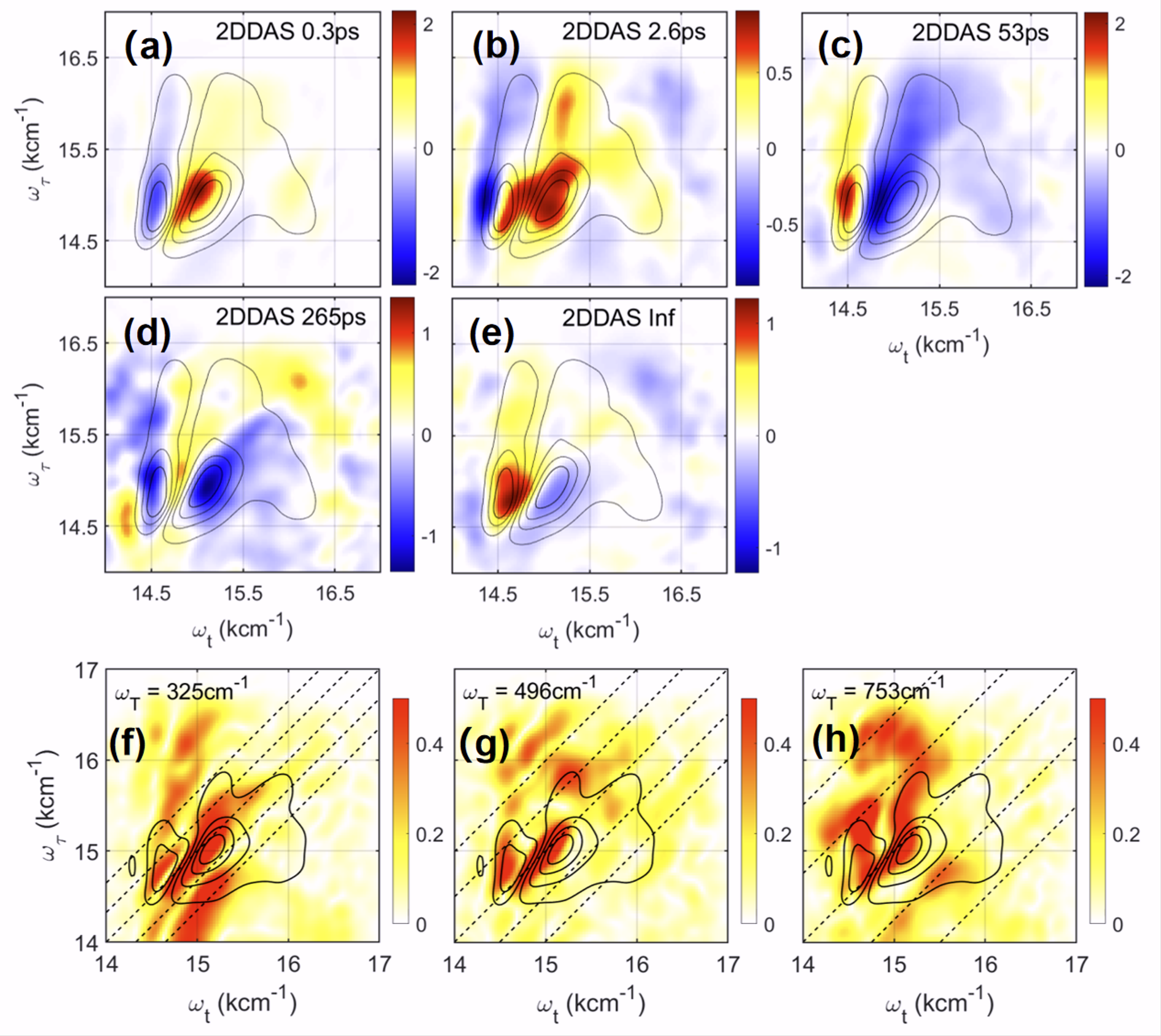} 
\caption{\label{fig:Fig3} Decay-associated spectral components from global analysis of PSI at 8 K. (a) The 0.3 ps component captures ultrafast downhill transfer within the antenna. (b) The 2.6 ps component reflects further equilibration and depletion of bulk chlorophyll states along the diagonal. (c) The 53 ps contribution highlights population transfer into red pools across both core and LHCR. (d) The 265 ps component indicates progressive accumulation of long-lived excitations in these sinks. (e) The infinite component displays signifying excitons stabilised in the lowest-energy states. (f-h) Fourier-transformed oscillatory maps reveal vibrational coherences with characteristic frequencies of 325, 496, and 753 cm$^{-1}$.  } 
\end{center}
\end{figure}

Application of the global fitting procedure resolved five decay-associated spectral (2DDAS) components, spanning from 0.3 ps to an effectively non-decaying contribution. These maps are shown in  Fig.\ \ref{fig:Fig3}(a-e), with contours of the original 2DES spectra overlaid to highlight the involved regions. The fastest component (0.3 ps; Fig.\ \ref{fig:Fig3}(a)) is characterised by a strong positive-negative pair centred around  ($\omega_{\tau}$, $\omega_{t}$) = (15,500, 14,500) cm$^{-1}$, which we attribute to ultrafast downhill energy transfer from higher-lying antenna pigments into lower-energy states. This sub-picosecond equilibration reflects the rapid redistribution within strongly coupled chlorophyll domains. The following 2.6 ps component (Fig.\ \ref{fig:Fig3}(b)) displays a diagonal bleach feature centred near 15,000 cm$^{-1}$, signifying depletion of bulk antenna states. On longer timescales, the 53 ps component (Fig.\ \ref{fig:Fig3}(c)) exhibits complementary blue signatures at the low-energy side, consistent with population arrival into red pools of both the LHCR and core. The 265 ps contribution (Fig.\ \ref{fig:Fig3}(d)) reveals more pronounced accumulation in these sinks, highlighting their role as long-lived reservoirs. Finally, the effectively infinite component (Fig.\ \ref{fig:Fig3}(e)) retains a persistent red bleach in the ($\omega_{\tau}$, $\omega_{t}$) $\approx$ (14,500–15,000) cm$^{-1}$ region, corresponding to excitations localised in the terminal red chlorophyll states. This enduring signal represents the stabilised precursors to charge separation.

Beyond population dynamics, the 2DES data reveal oscillatory features indicative of vibrational coherence. Such oscillations observed in 2DES studies of different photosynthetic complexes were initially proposed as signatures of coherent excitonic superpositions, implying that wave-like transport might augment classical hopping \cite{Ref21}. Subsequent experimental and theoretical studies, however, demonstrated that these signals predominantly arise from vibrational coherences \cite{Ref22, Ref23, Ref24, Ref25}. To probe these contributions in our system, we analyzed residuals from global fitting and constructed frequency-resolved vibrational maps. The resulting maps (Fig.\ \ref{fig:Fig3}(f-h)) exhibit oscillations at 325, 496, and 753 cm$^{-1}$, aligned along the antidiagonal, consistent with vibrational progressions coupled to the excitonic manifold. Additional maps in the SI confirm that multiple coherent modes persist across PSI antenna and reaction center regions. Based on prior spectroscopic assignments, the 325 cm$^{-1}$ band reflects porphyrin-ring deformations and Mg-N bending in Chla \cite{Novoderezhkin2011}, the 496 cm$^{-1}$ mode corresponds to skeletal deformations and vinyl-macrocycle bending \cite{Dostal2016}, and the 753 cm$^{-1}$ band to C-C/C-N macrocycle stretching, long recognized as coherence carriers in PSI and LHC proteins \cite{Maiuri2018}. Their presence across diagonal and cross-peak regions suggests possible roles in both bulk antenna relaxation and feeding of red pools. The functional significance of these vibrational modes has not been validated in the present study and therefore merits focused investigation in future work.

\subsection{Temperature dependence of excitonic relaxation}

\begin{figure}[h!]
\begin{center}
\includegraphics[width=15.0cm]{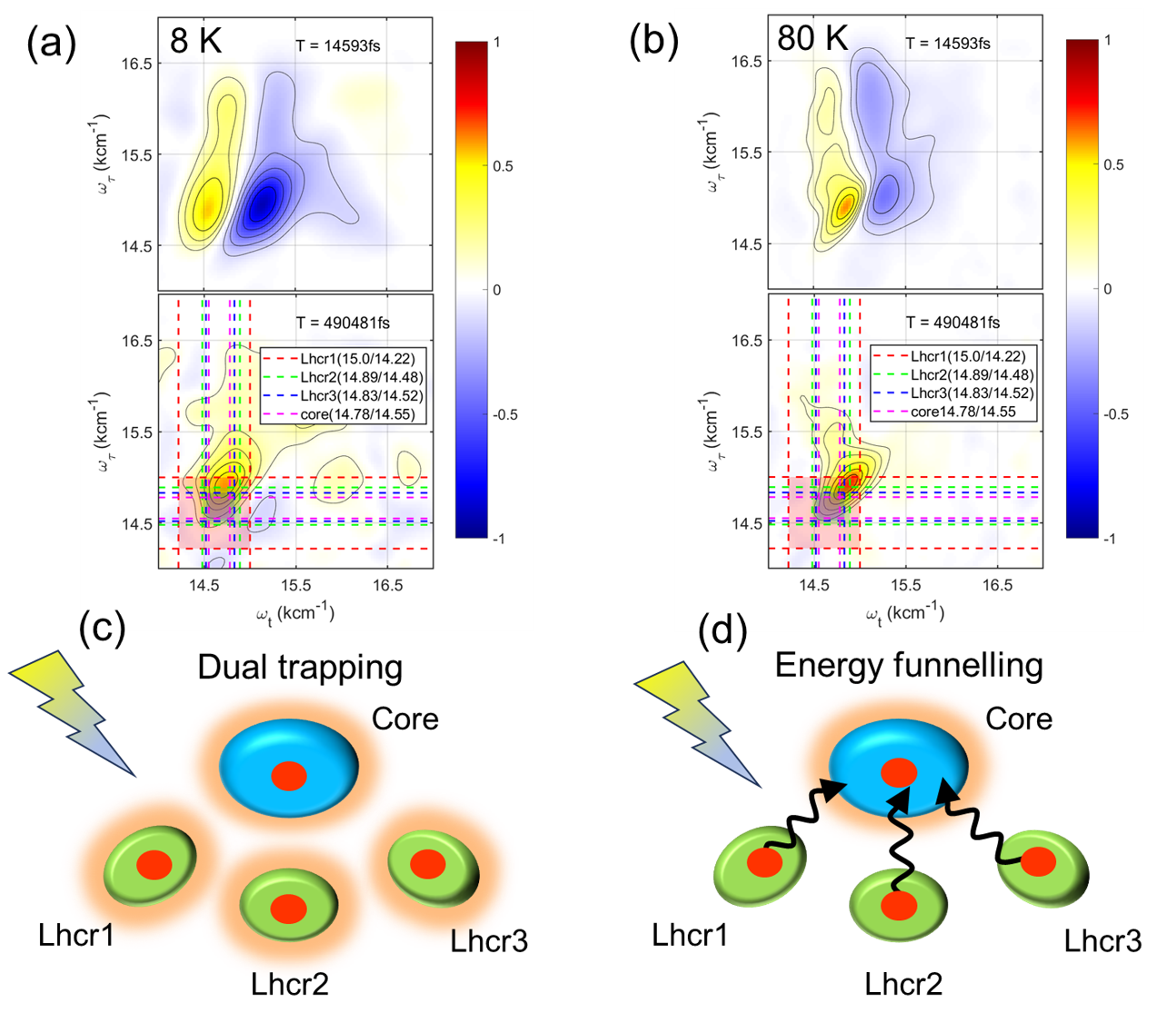}
\caption{\label{fig:Fig4}Temperature-dependent dynamics of PSI red pools. Two-dimensional spectra at 8 K (left) show broad ground-state bleach features reflecting heterogeneous population of multiple red pools in LHCR and the core. At 80 K (right), features sharpen and converge into the core region, consistent with thermally assisted funneling into dominant sinks. Dashed lines mark the site energies of Lhcr1-3 and the core red pigments. Schematics below, (c) and (d), illustrate this transition from distributed to thermally guided trapping pathways. } 
\end{center}
\end{figure}

To investigate how temperature modulates excitonic relaxation dynamics, we directly compared 2DES measurements of PSI at 8 K and 80 K (Fig.\ \ref{fig:Fig4}). Full 80 K data sets are available in SI as Figs. S4-S7. At early waiting times (T $\sim$15 ps), both temperatures show characteristic GSB and SE features near the diagonal, accompanied by negative ESA below. Notably, the 2DES spectra reveal robust downhill energy transfer from high-energy Chls into red-shifted acceptors. However, striking differences emerge at later time delays (T $\sim$ 500 ps). At 8 K, GSB features below 14,600 cm$^{-1}$ persist but are spectrally broad and spatially dispersed across Lhcr1 $\sim$(15,000/14,220 cm$^{-1}$), Lhcr2 $\sim$(14,890/14,480 cm$^{-1}$), Lhcr3 $\sim$(14,830/14,520 cm$^{-1}$), and the core red pool $\sim$(14,780/14,550 cm$^{-1}$), indicating heterogeneous exciton trapping across multiple sites. In contrast, at 80 K, GSB features sharpen markedly and converge more strongly within the core red region, suggesting selective thermal funneling. This interpretation is supported by 2DDAS analysis (Fig.\ \ref{fig:Fig3}(a-e) and SI Fig. S5), which resolves comparable kinetic components at both temperatures: sub-ps equilibration (0.3 - 0.8 ps), fast bulk-state depletion (2.6 - 4 ps), red-pool feeding (18 - 53 ps), and long-lived terminal population ($\ge$265-537 ps). Yet, the spectral amplitudes of these components differ: 8 K traces show broader decay features and greater spectral congestion, while 80 K signals display narrower, better-resolved transitions into the core sink. These findings indicate that the temperature does not introduce new kinetic phases but redistributes exciton populations: the slower red-pool signals become spectrally sharper at 80 K, reflecting more convergent trapping compared to the heterogeneous sink occupation seen at 8 K. Thus, while the overall energy-transfer pathways are conserved, thermal energy reshapes the exciton trapping landscape by biasing population flow towards specific red-chlorophyll sites.

\subsection{Theoretical modelling of excitonic population dynamics} 

\begin{figure}[h!]
\begin{center}
\includegraphics[width=15.0cm]{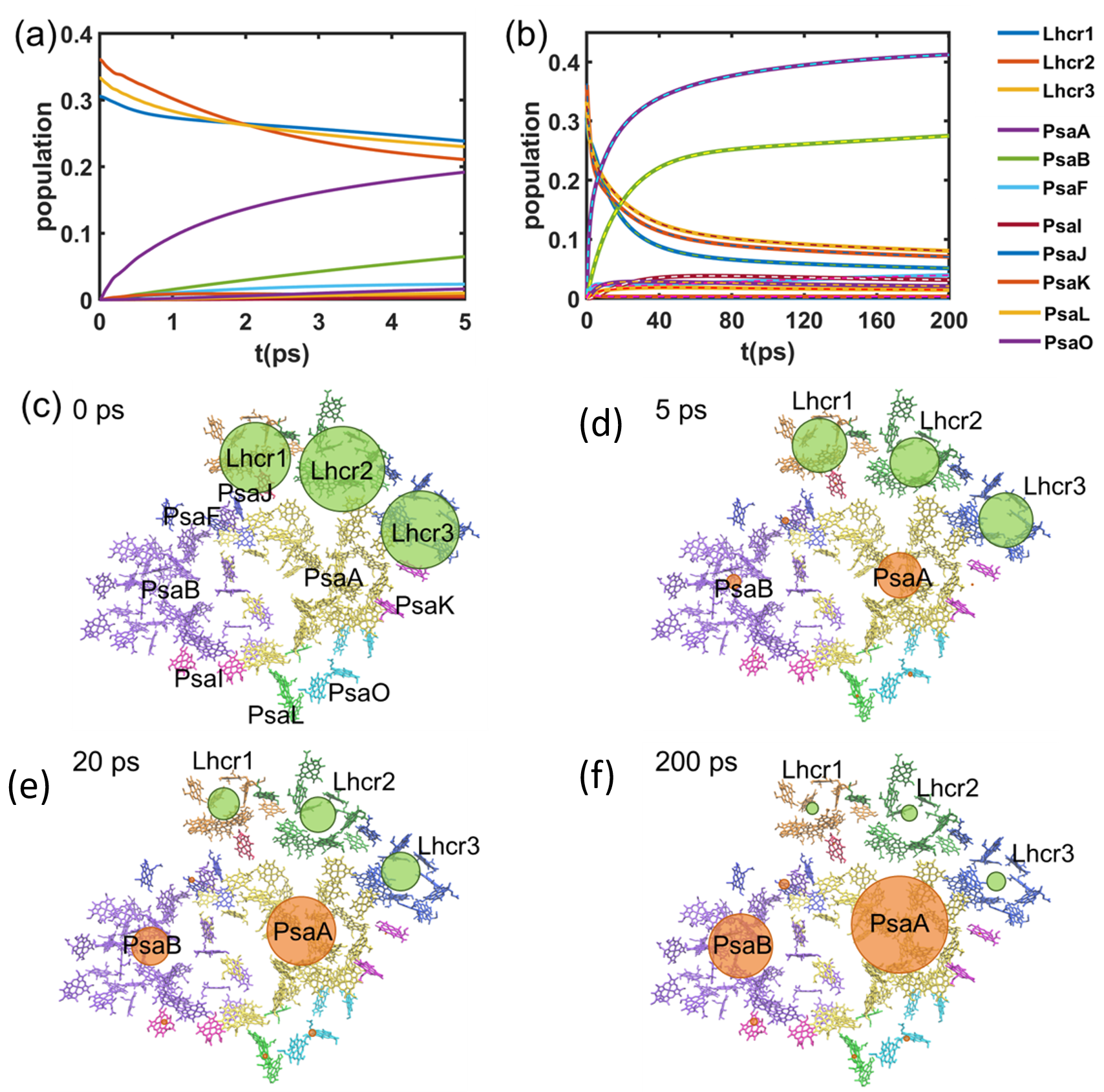}
\caption{\label{fig:Fig5} Theoretical modeling of exciton dynamics in PSI. (a, b) Population dynamics simulated with a time-nonlocal quantum master equation using Hamiltonians derived from {\em ab-initio} site energies and dipole-dipole couplings. Calculations reveal rapid equilibration within LHCRs and the PSI core (sub-ps to few ps), followed by slower population of low-energy red states. Black dashed curves indicate kinetic fits. (c-f) Structural maps illustrate population flow from antenna (Lhcr1-3) into distributed sinks within the core (PsaA/B) over 0-200 ps.} 
\end{center}
\end{figure}

To complement the experimental 2DES results and provide a microscopic picture of the exciton relaxation in PSI, we performed quantum-dynamical simulations based on an atomistic excitonic Hamiltonian. The site energies of all Chl-pigments were obtained from {\em ab-initio} electronic structure calculations, while excitonic couplings were evaluated within the dipole-dipole approximation. The full details of these calculations and the resulting Hamiltonian are provided in the SI. Together, these parameters define the electronic landscape across the PSI core and LHCR antenna complexes, enabling us to probe how strongly coupled pigments form delocalized excitonic states. The subsequent energy-transfer dynamics were computed within the framework of a time-nonlocal (TNL) quantum master equation. In this approach, each pigment is modeled as a two-level system coupled to a structured phonon bath that mimics protein-induced fluctuations. The bath was described as a set of harmonic oscillators with spectral densities chosen to reproduce both fast intramolecular vibrations and slower protein motions, as established in prior work on PSI and related complexes. This system-bath formalism captures both dissipative relaxation and coherent oscillatory contributions, making it particularly well-suited to interpret the experimentally observed vibrational coherences. Compared to simpler rate-equation models, the TNL method retains non-Markovian memory effects, allowing us to account for transient coherence and the role of vibronic mixing. Figure~\ref{fig:Fig5}(a) presents the simulated short-time population dynamics across PSI subunits, assuming uniform initial excitation of all three LHCRs. The calculations reveal rapid equilibration within LHCR units, occurring within the first picosecond, in close agreement with the experimentally resolved $\sim$100 fs - 1 ps components. On longer timescales (Fig.~\ref{fig:Fig5}b), excitations redistribute toward both the core (PsaA/B) and low-energy red pools located in LHCR and the core, consistent with the dual-sink architecture of \textit{C. merolae}  PSI. Multi-exponential fits to the simulated traces (dashed lines) yield characteristic timescales from sub-ps equilibration to $\sim$20 ps for red-pool feeding, quantitatively matching the decay-associated spectral components obtained from experiment. The extracted timescales and amplitudes are summarized in Tab.\ 1 in SI.


\section*{Discussions} 

Our results offer a coherent mechanistic picture of excitation-energy transfer in PSI from \textit{C. merolae}, a species that holds a key evolutionary position between cyanobacteria and the green lineage. By integrating cryogenic 2DES with room-temperature quantum-dynamical simulations, we resolve how energy is redistributed across pigments and clarify the functional role of red chlorophylls. A central observation in this work is the recognition that \textit{C. merolae} PSI employs a dual-sink arrangement, with low-energy pigments located both within the core and the LHCR antennae. Our temperature-dependent 2DES results provide critical perspective on the functional relevance of dual-sink trapping in \textit{C. merolae} PSI. Although this extremophilic red algae never experience cryogenic conditions, measurements at 8 K reveal how excitations distribute when thermal motion is essentially frozen (kT $\sim$5.5 cm$^{-1}$). Under these conditions, excitations remain in local minima, yielding broad ground-state bleach signals that directly expose the heterogeneity of red chlorophyll traps across both LHCR and the core.  At elevated cryogenic and physiological temperatures, however, thermal activation becomes significant: kT $\sim$55 cm$^{-1}$ at 80 K and $\sim$208 cm$^{-1}$ at 300 K are comparable to the energy gaps between antenna reds (e.g., Lhcr1 at $\sim$14 220 cm$^{-1}$) and core red chlorophylls ($\sim$14 550 cm$^{-1}$). Thus, even a modest increase in temperature enables excitations to overcome uphill gaps of 100 – 300 cm$^{-1}$ and converge into the core-associated sinks. Importantly, this does not require each exciton to carry the full energy difference; rather, thermal disorder in the protein matrix provides a distribution of phonon modes and entropic fluctuations that occasionally supply sufficient quanta to bridge the gap. In this way, dynamic sampling of vibrational environments ensures that even energetically 'trapped' excitations can be liberated and redirected toward the reaction center at physiologically relevant temperatures. Functionally, this means that under physiological conditions PSI behaves as a single effective funnel to P700, while the apparent ‘dual-sink’ behaviour is suppressed. Nonetheless, the existence of multiple red traps is not merely vestigial. They expand the spectral range of light harvesting and provide backup pathways under conditions where the RC is saturated or transiently closed, ensuring both robustness and efficiency in PSI energy conversion \cite{Joanna paper}. 

Taken together, these observations suggest that under physiological conditions, excitations are funneled more efficiently by exploiting the robustness provided by the distributed sinks. In other words, the architecture of \textit{C. merolae} PSI appears tuned for flexibility: under conditions of low disorder, multiple pathways are maintained, while at operational temperatures, efficient transfer is ensured. This interpretation is reinforced by our quantum-dynamical simulations at room temperature, which reproduce the experimental timescales and reveal how excitations partition between core and LHCR sinks. The theory indicates that equilibration is accelerated at physiological conditions. Overall, such a mechansim of dual trapping and selective funneling is absent in cyanobacteria, where a single core-localized red pool dictates dynamics, and in plants, where long-lived LHCI charge-transfer states dominate. \textit{C.\ merolae} thus represents a true evolutionary midpoint: retaining the efficiency of core-based trapping while introducing antenna-associated sinks that expand functional diversity. Altogether, these findings highlight how diversification of antenna subunits reshaped PSI during eukaryotic evolution. Rather than simply adding pigment density, optimized antenna expansion led to redistributed low energy sinks, creating new ways to balance efficiency, flexibility, and photoprotection. In this sense, \textit{C. merolae} PSI exemplifies how red pigments function as both spectral enhancers and kinetic regulators. Their dual role provides a unifying framework for understanding PSI evolution and offers a design principle that may inspire future artificial light-harvesting systems.


\section*{Conclusion} 

In this study, we have combined 2DES with theoretical modelling to disentangle the excitonic dynamics of PSI from \textit{C.\ merolae}, an organism occupying a pivotal evolutionary niche between cyanobacteria and plants. By undertaking measurements under cryogenic conditions, thermal disorder was sufficiently suppressed to resolve otherwise overlapping spectral features, permitting identification of discrete kinetic components and clarification of how excitation energy is channelled through both antenna and core pigments. Global analysis revealed five characteristic timescales, from sub-picoseond equilibration within LHCR subunits to tens-of-picoseconds population transfer into low-energy red pools. Crucially, the results demonstrate that \textit{C.\ merolae} PSI sustains dual excitonic sinks distributed across LHCR and the core, in contrast to the strictly core-localised red pools of cyanobacteria and the LHCI-dominated sinks of higher plants. Temperature-dependent measurements further illuminate the functional implications of this architecture. At 8 K, excitations remain heterogeneously distributed across multiple low-energy sites, directly exposing the inherent static disorder of the red pools. By 80 K, however, modest thermal activation (kT $\sim$55 cm$^{-1}$) is already sufficient to overcome some of the uphill gaps of 100 - 300 cm$^{-1}$ between antenna and core pigments, thereby biasing trapping into the core-associated sinks. At physiological temperatures, where kT approaches $\sim$200 cm$^{-1}$, entropic fluctuations of the protein–phonon environment ensure that such barriers are readily surmounted, producing a single effective funnel into the reaction centre. Thus, the dual-sink architecture is not vestigial, but represents a balanced evolutionary strategy: red chlorophylls act simultaneously as spectral enhancers, broadening the usable solar spectrum, and as latent kinetic regulators, providing redundancy under conditions of reduced thermal disorder or transient reaction-centre closure. More broadly, this work advances a coherent mechanistic framework that reconciles long-standing debates concerning the origin and role of PSI red states. It reveals how modifications to antenna–core connectivity shape the energy-transfer funnel, yielding both efficiency and resilience. These insights not only illuminate a critical evolutionary intermediate but also furnish design principles for artificial light-harvesting systems, where the deliberate introduction of multiple low-energy sinks may confer enhanced robustness without compromising overall photochemical yield.

%


\section*{Materials and Methods}
\subsection{Sample preparation.} 

The genetic strategy to introduce the modified His$_6$-tagged psaD construct (CMV144CT, 417 bp) into the URA locus of the \textit{Cyanidioschyzon merolae} M4 mutant was adapted from the protocol of Fujiwara {\em et al.} \cite{Joanna ref1}. The biochemical characterization of the recombinant PSI complex for assessment of purity, structural integrity, and photochemical activity of the biophotocatalyst, in comparison with the native PSI complex, are reported in Izzo {\em et al.} \cite{Joanna ref2}. The engineered {\em C.\ merolae} strain was cultivated in Allen 2 medium (pH 2.5) at 42 $^{\circ}$C under continuous white light illumination (90 $\mu$mol photons·m$^{-2}$·s$^{-1}$; Panasonic FL40SS-ENW/37) with gentle bubbling with 3-5\% CO$_2$ in air, as previously described in Ref.\ \cite{Joanna ref3} and \cite{Joanna ref4}. Cultures (9 L) were grown to an OD680 of 0.9-1.0 for isolation of thylakoids and PSI purification, as described in Haniewicz {\em et al.} \cite{Joanna ref4}. For thylakoids isolation, cells were disrupted on ice with 0.1 mm glass beads for 13 cycles (10 s `on', 4 min `off') in buffer A (10 mM CaCl$_{2}$, 5 mM MgCl$_{2}$, 25\% (w/v) glycerol, 40 mM MES-NaOH, pH 6.1) supplemented with DNase I (5 mg), RNase (10 $\mu$L), and a protease inhibitor cocktail (Thermo Fisher) (1 tablet per 50 mL). The homogenate was filtered (whatman paper), then centrifuged at 180,000 xg for 25 min at 4$^{\circ}$C to pellet thylakoids. Thylakoid membranes were washed three times with buffer A, resuspended to final Chl a concentration of 2-5 mg$\cdot$ mL$^{-1}$, then snap-frozen in liquid nitrogen for further use. 

The His$_6$-PsaD-PSI complex was purified, as reported in Izzo {\em et al.} \cite{Joanna ref2}. Briefly, the His-tagged complex was purified using immobilized metal affinity chromatography (IMAC). Thylakoids (1 mg Chl a$\cdot$mL$^{-1}$ were solubilized with 1\% (w/v) n-dodecyl-$\beta$-D-maltoside (DDM) in the dark on ice for 40 min. All the subsequent procedure were performed at 4$^{\circ}$C. The solubilized fraction was filtered and loaded onto a 1 mL HisTrap HP column (Cytiva), pre-equilibrated with 5 column volumes (CV) of the wash buffer (3 mM CaCl2, 0.03\% DDM, 25\% glycerol, 20 mM imidazole, 40 mM HEPES-NaOH, pH 8.0). After washing with 10 CV of the wash buffer, bound PSI complex was eluted using a linear imidazole gradient (0-1 M) in the same buffer. Eluted fractions were analyzed via UV-VIS absorbance spectroscopy (Shimadzu UV-VIS 1800) (Fig. SXA). Imidazole was removed by buffer exchange using Vivaspin-20 concentrators (100 kDa MWCO, Sartorius) at 4000xg, 4 for 15 min. Elution buffer was diluted 1:1 with wash buffer without imidazole and the exchange repeated two more times. 

\subsection{2D Electronic measurements with experimental conditions.} 

The details of the experimental setup follow previous reports from our group \cite{Ref25, Ref45}. In brief, two-dimensional electronic spectra were recorded using a diffractive optics-based, all-reflective 2D spectrometer providing phase stability of $\lambda/160$. Excitation pulses were generated by a home-built nonlinear optical parametric amplifier (NOPA), pumped by a commercial femtosecond laser system (Spectra Physics, Newport). The pulses were compressed to $\sim$12 fs using a combination of a deformable mirror (OKO Technologies, 19-channel) and a fused silica prism pair. Their temporal profile was characterized by frequency-resolved optical gating (FROG), with the traces analyzed using the commercial package FROG3 (Femtosecond Technologies). The resulting broadband spectrum had a $\sim$100 nm full width at half maximum (FWHM), centered at 620 nm, sufficient to cover the main electronic transitions into the lowest excited states. For 2DES measurements, three beams were focused onto the sample with a spot size of $\sim$130~$\mu$m, generating a photon-echo signal along the phase-matching direction. The emitted signal was collected using a Sciencetech 9055F spectrometer coupled to a CCD linear array detector (Entwicklungsb{\"u}ro Stresing). For all experiments, the excitation pulse energy was attenuated to $\sim$10 nJ at a 1 kHz repetition rate. Phasing of the TG data was achieved using the “invariant theorem” procedure described in Ref.\ \citeonline{David2001}.

\subsection{Theoretical calculations}

We employed the {\em ab-initio} method to optimize the molecular structure of PSI complex. We then employed the TDDFT method to calculate the electronic ground and excited states. Before that, the molecular structures of subunits were also examined by frequency calculations at ground state, which allowed the molecular structure of pigments and associated protein matrix at their stable structure. Moreover, the excitonic couplings between pigments inside antenna complexes and also in core complexes have been investigated by dipole-dipole interactions. The electronic excited structures of pigments have been modeled as two-level model: electronic ground and excited states. The system Hamiltonian can be constructed with the optimal site energies and the off-diagonal elements, excitonic couplings. With these parameters, the system Hamiltonian can be written as 
\begin{equation}
\label{eq:system Ham} 
H_S = \sum_{i=1}^N  \ket{e_i}\varepsilon_i\bra{e_i} + \sum_{i=1}^N\sum_{\substack{\\ i \neq j}}  \ket{e_i}V_{ij}\bra{e_j}, 
\end{equation} 
where $\varepsilon_i$ is the site energy (energy gap between ground and first electronic excited state) of i-th pigment. $V_{ij}$ is the excitonic coupling between i- and j-th pigments. In PSI complex, we have sorted more than 132 pigments in 11 complexes. Thus, we have N=132. We also employed the system-bath model for the calculations of energy transfer. For this, the noisy environment can be modeled as infinity number of harmonic oscillators, thus, we have 
\begin{equation}
\label{eq:bath Ham} 
H_{\mathrm{env}} = \sum_{i=1}^{N} \sum_{k=1}^{N_b^i} \left( \frac{p_{ik}^2}{2} + \frac{1}{2} \omega_{ik} x_{ik}^2 \right).	
\end{equation}
Here, N$_b^i$ is the number of bath modes coupled to pigment i, while x$_{ik}$ and p$_{ik}$ are the mass-weighted position and momentum of the k-th harmonic oscillator mode with frequency $\omega_{ik}$. The system-bath coupling can be modeled as the interactions of electronic states to the coordinates of harmonic oscillators, which yields 
\begin{equation}
\label{eq:sys-bath Ham} 
H_{\mathrm{sb}} = \sum_{i} \ket{e_i}\bra{e_i} \sum_{k} c_{ik} x_{ik}, 
\end{equation} 
where c$_{ik}$ represents the  coupling between the i-th pigment and the k-th pigment. The bath spectral density is defined as 
\begin{equation}
\label{eq:SD} 
J_i(\omega) = \pi \sum_j \frac{c_{ij}^2}{2m_{ij}\omega_{ij}} \delta(\omega - \omega_{ij}) =  \gamma_{0}  \omega  e^{-\omega / \omega_{c}}, 
\end{equation}
Here, $\gamma_0$ is the coupling strength, $\omega_c$ is the cut-off frequency of the bath. The influence of the bath can be fully described by its bath spectral density, which takes the form shown in Eq.\ \ref{eq:SD} with parameters $\gamma_{0}$ = 0.7 cm$^{-1}$ and $\omega_{c}$ = 100 cm$^{-1}$. Moreover, the obtained site energies of pigments and the associated excitonic couplings between them have been described in the SI. The population dynamics has been calculated by TNL method, the detailed description of it has been written in the SI. With the optimal parameters and also the calculations, we obtained the time resolved population dynamics.

%
\begin{addendum}
\item This work was supported by the National Key Research and Development Program of China (Grant No.\ 2024YFA1409800), NSFC Grant with No.\ 12274247, Yongjiang talents program with No.\ 2022A-094-G and 2023A-158-G, Ningbo International Science and Technology Cooperation with No.\ 2023H009, `Lixue+' Innovation Leading Project and the foundation of national excellent young scientist. The Next Generation Chemistry theme at the Rosalind Franklin Institute is supported by the EPSRC (V011359/1 (P)) (A.J.). This work was funded by the National Science Center, Poland (Solar-driven chemistry grant no.\ 2022/04/Y/ST4/00107 to JK). RJDM is supported by the Natural Sciences and Engineering Research Council of Canada. 

\item[Supporting information] The details of model Hamiltonian and system-bath model, {\em ab-initio} calculations of site energies and excitonic couplings are described. Moreover, the TNL quantum master equation, global fitting approach and detailed data treatment of fitting procedures, residuals have been introduced as well. The 2DES and resulted data of PSI complex at different temperature has also been described. 

\item[Competing Interests] The authors declare that they have no competing financial interests. 

\item[Correspondence] Correspondence of paper should be addressed to R.J.D.M. ~(dmiller@lphys2.chem.utoronto.ca), J.K. ~(j.kargul@cent.uw.edu.pl), F.Z. ~(zhengfulu@nbu.edu.cn), A.J. ~(Ajay.Jha@rfi.ac.uk) and H.-G.D. ~(email: duanhongguang@nbu.edu.cn).  

\end{addendum}
%
\newpage

\newpage

\newpage

\newpage

\newpage

\end{document}